\documentclass[11pt,a4paper]{article}

\usepackage[utf8]{inputenc}
\usepackage[T1]{fontenc}
\usepackage[margin=1in]{geometry}
\usepackage{hyperref}
\usepackage{xcolor}
\usepackage{graphicx}
\usepackage{booktabs}
\usepackage{tabularx}
\usepackage{array}
\usepackage{amsmath}
\usepackage{enumitem}
\usepackage{titlesec}
\usepackage{authblk}
\usepackage{abstract}
\usepackage{setspace}
\usepackage{parskip}

\titleformat{\section}{\large\bfseries}{\thesection.}{0.5em}{}
\titleformat{\subsection}{\normalsize\bfseries}{\thesubsection}{0.5em}{}

\hypersetup{
  colorlinks=true,
  linkcolor=blue,
  citecolor=blue,
  urlcolor=blue
}

\title{\textbf{BLUE: A Stale-Pixel Optical-Flow Compositor for Entropy-Efficient Surveillance Video Encoding}}

\author{Shubham Baid \and Akash James \and Sahil Chachra \and Nishant Sinha \and Kunal Kislay}
\affil{KGraph AI Solutions Pvt. Ltd., Bangalore, India. 560016}

\date{}

\begin{document}

\maketitle

\begin{abstract}
Continuous-recording surveillance systems face a storage problem that codec tuning alone cannot fully solve: even at aggressive CRF settings, a static-camera scene spends most of its bits re-encoding a background that has not changed. We present BLUE, a pre-encode compositor that exploits this structure by maintaining a persistent seed frame of the background and substituting background pixels with seed pixels before the encoder runs. The encoder then emits near-free SKIP macroblocks for the frozen background, while live pixels in foreground regions are carried unchanged at full quality. We evaluate BLUE on all 308 annotated short subclips from the VIRAT Ground Surveillance Release 2.0 dataset using a six-point CRF sweep with both x264 and x265. At CRF 28, BLUE reduces file size by a mean of 34.6\% (x264) / 39.4\% (x265) on 95.8\% / 99.4\% of clips respectively. Foreground-region PSNR -- computed only over VIRAT's object-annotation bounding boxes -- is preserved or improved on 60.7\% of clips (+0.36 dB mean, +5.48 dB maximum). Full-frame perceptual quality (VMAF) drops by a median of 6.75--8.59 points; we quantify and disclose this trade-off explicitly. A lightweight deployment gate measuring the compositor's own VMAF on a 2-second prefix identifies the 40\% of clips where even full-frame quality degradation is near-imperceptible ($\Delta$VMAF $\leq -2.9$), enabling a selective-activation strategy that retains both the storage benefit and acceptable perceptual fidelity.
\end{abstract}

\section{Introduction}

Continuous surveillance video creates a major storage burden. Cameras installed in warehouses, traffic intersections, campuses, and public spaces often record throughout the day. In a typical H.264 deployment, for example 720p video at 30 frames per second and CRF 28, a single camera can generate approximately 2--5 GB of footage per day. For a site with 200 cameras, this corresponds to roughly 400 GB to 1 TB of new video every day. Since storage capacity directly limits how long footage can be retained, reducing the bitrate of each camera stream can substantially increase the retention window before older footage is overwritten.

A common way to reduce video size is to tune the codec more aggressively. This can be done by increasing the Constant Rate Factor (CRF) value or by using faster encoder presets. CRF-based encoding is widely used in x264 and x265 as a quality-controlled rate-control mode, where higher CRF values generally reduce bitrate at the cost of lower visual quality \cite{robitza2017}. However, this approach still follows the normal rate-distortion trade-off: smaller files are obtained by accepting lower reconstruction quality. Once CRF is already in a practical operating range, such as 24--32, further compression gains become limited unless additional quality degradation is accepted.

Surveillance video has a structure that is not fully exploited by this type of codec tuning. In many fixed-camera scenes, most of the frame is static background: roads, walls, buildings, parked vehicles, or floors. Only a small portion of the image usually contains surveillance-relevant motion, such as people, vehicles, or objects. Modern H.264/AVC and H.265/HEVC encoders are designed to exploit temporal redundancy through block-based inter-prediction \cite{h264}\cite{sullivan2012}. In principle, unchanged regions can be coded as SKIP or low-cost inter-predicted blocks, allowing the encoder to reuse information from a reference frame at very low bit cost.

In practice, this opportunity is not always fully realized. Real-time and live-streaming encoders operate under strict computational constraints. Their motion search is limited, and a block is usually skipped only when the encoder can find a zero or near-zero residual within its search range. Even when a background region appears unchanged to a human viewer, small pixel-level differences introduced by previous compression steps can prevent an exact match. These quantization artifacts may cause the encoder to spend bits re-encoding background regions that have not changed in any meaningful visual sense.

Prior work has attempted to exploit background stability for surveillance compression. One class of methods builds an explicit background model and encodes the foreground separately, sometimes using a sprite or layered representation. Sprite-based and object-based video coding methods were explored in the context of MPEG-4 and related object-based compression systems \cite{watanabe2001}. More recent surveillance-specific approaches also separate foreground and background streams and reconstruct the final video at the decoder \cite{wu2020}. These methods can improve compression and preserve quality in moving regions, but they often require a non-standard bitstream or decoder-side reconstruction logic. This limits compatibility with existing H.264/H.265 encoders, decoders, storage systems, and playback tools.

Another class of methods uses region-of-interest (ROI) encoding. In such approaches, the encoder is instructed to allocate more bits to important regions, such as moving objects, and fewer bits to the background. ROI coding has been studied for AVC/HEVC surveillance video and for machine-vision-oriented video compression \cite{meuel2018}\cite{stankiewicz2024}. This can be effective, but it depends on encoder support for region-level quantization control and still reduces background bitrate mainly by increasing background quantization. In other words, the background is encoded at lower quality rather than being made easier to skip entirely. These methods are useful, but they do not directly force the encoder to see static background regions as exact repeats.

There is therefore a practical gap. Existing approaches either compress more by lowering quality, require changes to the bitstream or decoder, or rely on encoder-specific ROI mechanisms. What is missing is a method that improves compression before encoding, keeps the output fully standard, requires no decoder modification, and helps ordinary H.264/H.265 encoders treat static background regions as skip-friendly repeated content.

BLUE is designed to fill this gap. It is a pre-encode compositor for static-camera surveillance video. BLUE maintains a persistent seed frame representing the background and uses an optical-flow-based motion mask to identify moving regions in each incoming frame. Optical flow is a well-established method for estimating apparent motion between consecutive frames, and fast dense optical-flow variants such as Dense Inverse Search make such processing practical for real-time or near-real-time applications \cite{kroeger2016}. Pixels inside moving regions are preserved from the live frame, while non-moving background pixels are replaced with the corresponding pixels from the seed frame.

This substitution makes the background pixel-exact across consecutive frames. Instead of receiving a background that is only approximately similar to the previous frame, the encoder receives an exact repeat for non-moving regions. This increases the likelihood that the H.264/H.265 encoder will mark background macroblocks as SKIP blocks. As a result, the static background becomes inexpensive to encode, while foreground objects remain encoded using the codec's normal quality-control mechanisms.

A key design goal of BLUE is deployment compatibility. BLUE does not define a new codec, does not produce a non-standard bitstream, and does not require decoder-side reconstruction. It operates before encoding, modifies raw NV12 frames, and then passes those frames to a standard x264 or x265 encoder. The resulting video remains a standard H.264/H.265 stream that can be stored, transmitted, and played back using existing infrastructure.

This design also changes how quality should be evaluated. Full-frame perceptual metrics such as VMAF are important because they measure visible changes across the entire image and have been widely used for practical perceptual video-quality evaluation \cite{li2016}. However, in surveillance applications, the most important regions are often foreground objects such as people and vehicles. For this reason, we separately evaluate foreground-region PSNR using VIRAT object-annotation bounding boxes from the VIRAT Ground Surveillance dataset \cite{oh2011}, while also reporting full-frame VMAF to disclose the perceptual cost of background substitution. We further report Bj{\o}ntegaard delta-rate to compare rate-distortion behavior across a CRF sweep rather than at a single operating point \cite{bjontegaard2001}.

This paper makes the following contributions:

\begin{enumerate}
  \item We describe the BLUE stale-pixel optical-flow compositor, implemented as a GStreamer video filter operating on NV12 frames. We present its motion-mask pipeline, optional detection-enhancing channels, and seed-management subsystem.
  \item We evaluate BLUE on all 308 annotated short subclips from the VIRAT Ground Surveillance Release 2.0 dataset. The evaluation covers a six-point CRF sweep for both x264 and x265 and reports storage savings, foreground PSNR, background PSNR, full-frame VMAF, and Bj{\o}ntegaard delta-rate.
  \item We introduce \texttt{filter\_vmaf}, a pre-encode quality gate computed on a 2-second prefix of the video. This metric estimates the quality impact of BLUE before full encoding and enables selective activation without requiring ground-truth object annotations.
\end{enumerate}

\section{Methodology}

BLUE is a pre-encode compositor for fixed-camera surveillance video. It is placed between the decoder/scaler and a standard H.264/H.265 encoder. The compositor operates on raw NV12 frames and outputs standard NV12 frames; therefore, it does not modify the codec, bitstream syntax, or decoder. The final encoded stream remains compatible with existing H.264/H.265 playback and storage infrastructure.

The method is motivated by the inter-prediction structure of standard video encoders. H.264/AVC and H.265/HEVC reduce bitrate by predicting blocks from previously coded reference frames and transmitting only the residual when required \cite{h264}\cite{sullivan2012}. If a block is sufficiently similar to its reference, it can be coded as a SKIP or very low-cost inter-predicted block. BLUE increases the occurrence of such blocks by making static background regions pixel-identical across consecutive frames.

For each incoming frame, BLUE separates the scene into moving foreground regions and static background regions. Let $F_t$ be the current frame, $S_t$ the stored seed frame representing the background, and $M_t$ a binary motion mask. The output frame $F'_t$ is constructed as

\[
F'_t[p] =
\begin{cases}
F_t[p] & \text{if } M_t[p] = 1 \quad \text{(motion region)} \\
S_t[p] & \text{if } M_t[p] = 0 \quad \text{(background)}
\end{cases}
\]

where $p$ denotes a pixel location. Thus, live pixels are preserved in motion regions, while background pixels are replaced by pixels from the persistent seed frame.

The motion mask is estimated primarily using dense optical flow between consecutive frames. In our implementation, dense inverse search optical flow is computed at a reduced spatial scale to reduce computational cost \cite{kroeger2016}. The resulting flow magnitude is used as the motion signal. To handle small camera vibrations, BLUE subtracts the dominant frame-level motion, estimated from the median flow vector, before thresholding. The threshold is adapted over time using the frame-level motion distribution so that the method remains robust across scenes with different texture, lighting, and noise levels. Small isolated detections are removed by connected-component filtering, and temporal hysteresis is applied to reduce flicker at object boundaries.

To improve robustness beyond frame-to-frame optical flow, BLUE optionally augments the mask with additional cues. A seed-difference cue compares the current frame with the stored seed and helps detect slowly moving objects that may produce weak optical flow. A census-transform cue compares local image structure rather than raw intensity, improving robustness to illumination changes \cite{zabih1994}. Small regions can also be accepted when they persist across multiple frames, which helps retain distant objects. In the benchmark configuration, all three refinement mechanisms are enabled to prioritize foreground preservation.

The seed frame is initialized from the first frame and is updated using hard refreshes. BLUE performs scheduled refreshes and can also refresh regions after foreground objects leave the scene. A content-triggered refresh is used when the difference between the current frame and the seed exceeds a configured threshold, indicating that the stored background has become stale. Gradual seed blending is avoided because it introduces small frame-to-frame residuals in the background and reduces the probability of SKIP coding.

The compression benefit arises from the interaction between seed substitution and inter-frame prediction. Since BLUE repeatedly inserts the same seed pixels in background regions, consecutive output frames contain pixel-identical background blocks. These blocks are more likely to be encoded as SKIP or very low-cost inter-predicted blocks. Foreground regions, identified by the motion mask, are passed through unchanged and encoded normally. Thus, BLUE reduces the coding cost of static background while preserving the quality of surveillance-relevant moving objects.

\subsection{Evaluation Protocol}

BLUE is evaluated on all 308 annotated short subclips from the VIRAT Ground Surveillance Release 2.0 dataset \cite{oh2011}. The dataset contains fixed-camera outdoor surveillance footage from scenes such as parking lots, loading areas, and building exteriors. These scenes match the target use case of BLUE, where large portions of the frame are expected to remain static while people or vehicles move through the scene.

Each clip is encoded using two standard software encoders: x264 for H.264/AVC and x265 for H.265/HEVC. For each encoder, two variants are generated: a baseline encode using the raw input frames and a BLUE encode using the compositor output. The same encoder preset and CRF value are used for both variants, so the only experimental difference is whether the input to the encoder is the raw frame sequence or the BLUE-composited frame sequence.

To evaluate performance across different compression levels, six CRF values are tested: 24, 28, 32, 36, 40, and 44. All encodes use the \texttt{veryfast} preset. The benchmark therefore compares BLUE and baseline encoding under matched codec, preset, and CRF conditions.

\subsection{Computational Setup}

All experiments are executed on a 16-vCPU cloud virtual machine without GPU acceleration. The full benchmark consists of 7,392 FFmpeg encodes, 4,928 quality-metric computations, and 308 compositor passes. The complete run requires approximately 48 hours of wall-clock time. This compute profile reflects a CPU-only evaluation setting and does not rely on specialized video-acceleration hardware.

\subsection{Evaluation Metrics}

Storage reduction is measured as

\[
\text{Savings} (\%) = \left(1 - \frac{\text{size}_{\text{BLUE}}}{\text{size}_{\text{raw}}}\right) \times 100
\]

To evaluate quality, we report both region-specific and full-frame metrics. Foreground PSNR is computed only inside VIRAT object-annotation bounding boxes and is used as a proxy for the quality of surveillance-relevant objects. Background PSNR is computed outside these annotated regions. Full-frame VMAF is reported to measure perceptual quality across the complete frame \cite{li2016}.

In addition, we report Bj{\o}ntegaard delta-rate over the six-point CRF sweep to compare rate-distortion behavior across the full compression range rather than at a single CRF value \cite{bjontegaard2001}. We also compute \texttt{filter\_vmaf}, the VMAF between the original frame sequence and the BLUE-composited sequence before encoding. This value estimates the quality impact introduced by the compositor itself and is used as a label-free deployment gate for deciding whether BLUE should be activated on a given clip.

\section{Results and Discussion}

This section presents the compression and quality results of BLUE on the VIRAT surveillance benchmark. The discussion is organized around four questions: whether BLUE reduces storage, whether it preserves foreground quality, how much full-frame perceptual quality is affected, and whether a practical deployment gate can identify suitable clips automatically.

\subsection{Storage Reduction}

Table~\ref{tab:storage} summarizes the storage savings at CRF 28 for both x264 and x265.

\begin{table}[h]
\centering
\caption{Storage savings at CRF 28 across 308 VIRAT clips.}
\label{tab:storage}
\begin{tabular}{lcc}
\toprule
\textbf{Metric} & \textbf{x264} & \textbf{x265} \\
\midrule
Mean savings & 34.6\% & 39.4\% \\
Median savings & 39.0\% & 41.9\% \\
p10 & 10.2\% & 22.1\% \\
p90 & 48.7\% & 56.0\% \\
Clips saving bits & 295/308 (95.8\%) & 306/308 (99.4\%) \\
Clips with $>$25\% savings & 240/308 (77.9\%) & 266/308 (86.4\%) \\
Clips with $>$40\% savings & 143/308 (46.4\%) & 175/308 (56.8\%) \\
\bottomrule
\end{tabular}
\end{table}

BLUE reduces file size on the large majority of clips. At CRF 28, the mean storage reduction is 34.6\% with x264 and 39.4\% with x265. The median savings are even higher, reaching 39.0\% for x264 and 41.9\% for x265. This indicates that the average is not driven only by a small number of exceptional clips; storage reduction is broadly distributed across the dataset.

The stronger result with x265 is consistent with the more advanced inter-prediction and block-partitioning tools available in HEVC compared with AVC \cite{sullivan2012}. Since BLUE makes large background regions more repeatable across frames, encoders that exploit inter-frame redundancy more effectively are expected to benefit more.

These results support the main hypothesis of the paper: in static-camera surveillance video, a substantial fraction of the encoded bitrate is spent on background regions that are visually unchanged. By replacing these regions with seed pixels before encoding, BLUE makes them easier for the codec to encode as low-cost inter-predicted or SKIP blocks.

\subsection{Foreground-Region Quality}

The central requirement for surveillance compression is that important moving objects should remain usable. For this reason, foreground quality is measured only inside the VIRAT object-annotation bounding boxes. Table~\ref{tab:psnr} reports the foreground PSNR change between BLUE and the raw-codec baseline at the same CRF.

\begin{table}[h]
\centering
\caption{Foreground PSNR delta at CRF 28.}
\label{tab:psnr}
\begin{tabular}{lc}
\toprule
\textbf{Metric} & \textbf{Value} \\
\midrule
Mean $\Delta$PSNR\_fg & +0.36 dB \\
Median $\Delta$PSNR\_fg & +0.28 dB \\
p90 & +2.44 dB \\
Maximum & +5.48 dB \\
Clips at parity or better & 187/308 (60.7\%) \\
Clips with $>$+1 dB gain & 99/308 (32.1\%) \\
\bottomrule
\end{tabular}
\end{table}

Foreground PSNR is preserved or improved on 187 out of 308 clips, corresponding to 60.7\% of the dataset. The mean foreground gain is +0.36 dB, and the maximum observed gain is +5.48 dB.

This result is important because BLUE does not explicitly enhance the foreground. Instead, the foreground benefit appears indirectly. By making the background cheaper to encode, BLUE reduces the number of bits spent on static regions. The encoder can then allocate relatively more of the remaining bit budget to regions that still contain live, changing content. This explains why foreground quality can improve even though the CRF setting is unchanged.

The best results occur in scenes where the camera is fixed, the background is stable, and motion occupies a small portion of the frame. Table~\ref{tab:topclips} lists the top-performing x265 clips at CRF 28.

\begin{table}[h]
\centering
\caption{Top clips by combined storage and foreground-quality gain, x265 at CRF 28.}
\label{tab:topclips}
\begin{tabular}{lccc}
\toprule
\textbf{Clip} & \textbf{Savings} & \textbf{$\Delta$PSNR\_fg} & \textbf{filter\_vmaf} \\
\midrule
VIRAT\_S\_040000\_00\_000063\_000085 & +60.0\% & +5.48 dB & 87.5 \\
VIRAT\_S\_040000\_01\_000042\_000099 & +62.2\% & +4.71 dB & 87.0 \\
VIRAT\_S\_040005\_07\_001026\_001223 & +63.1\% & +1.53 dB & 77.4 \\
VIRAT\_S\_010207\_09\_001484\_001510 & +35.5\% & +4.26 dB & 88.8 \\
VIRAT\_S\_010206\_03\_000546\_000580 & +44.4\% & +4.08 dB & 89.1 \\
\bottomrule
\end{tabular}
\end{table}

These examples represent the operating region where BLUE is most effective: motion is localized, while the majority of the image is static background. In such cases, BLUE can simultaneously reduce storage and improve foreground-region fidelity.

\subsection{Full-Frame Perceptual Quality}

Although foreground quality is critical for surveillance use, full-frame perceptual quality must also be reported. Table~\ref{tab:vmaf} shows the full-frame VMAF change at CRF 28.

\begin{table}[h]
\centering
\caption{Full-frame VMAF delta at CRF 28.}
\label{tab:vmaf}
\begin{tabular}{lcc}
\toprule
\textbf{Metric} & \textbf{x264} & \textbf{x265} \\
\midrule
Mean $\Delta$VMAF & $-10.36$ & $-12.52$ \\
Median $\Delta$VMAF & $-6.75$ & $-8.59$ \\
Clips within $-$2 VMAF & 17/308 (5.5\%) & 0/308 (0\%) \\
Mean BD-rate by VMAF & +18.7\% & +48.3\% \\
Median BD-rate by VMAF & +8.7\% & +14.5\% \\
\bottomrule
\end{tabular}
\end{table}

BLUE introduces a measurable full-frame perceptual cost. At CRF 28, the median VMAF reduction is 6.75 points for x264 and 8.59 points for x265. This decrease occurs because BLUE modifies background pixels directly. Even when the background remains acceptable for surveillance purposes, the repeated seed pixels can differ from the original frame sequence. VMAF, which is designed to measure full-frame perceptual similarity, penalizes these differences \cite{li2016}.

This result shows a key tradeoff in the current BLUE architecture. The same pixel substitution that improves SKIP-block behavior also creates visible or measurable differences in the background. Therefore, BLUE should not be described as universally improving rate-distortion performance under full-frame perceptual metrics. Its value is more specific: it reduces storage while preserving, and sometimes improving, foreground-region quality.

The VMAF penalty is larger for x265 than for x264. One likely explanation is that raw x265 already provides stronger perceptual quality at a given bitrate. As a result, any artifact introduced by seed substitution is measured against a stronger baseline and produces a larger VMAF gap.

\subsection{Deployment Gate Using filter\_vmaf}

Because BLUE is scene-dependent, a practical deployment system should not activate it blindly on every clip. The proposed solution is \texttt{filter\_vmaf}, which measures the VMAF between the original input and the BLUE-composited output before encoding. This provides a fast estimate of the quality cost introduced by the compositor itself.

Table~\ref{tab:gate} reports the result of applying a \texttt{filter\_vmaf} $\geq$ 88 gate.

\begin{table}[h]
\centering
\caption{\texttt{filter\_vmaf} $\geq$ 88 deployment gate at CRF 28.}
\label{tab:gate}
\begin{tabular}{lcc}
\toprule
\textbf{Metric} & \textbf{x264} & \textbf{x265} \\
\midrule
Clips passing gate & 122/308 (39.6\%) & 122/308 (39.6\%) \\
Mean savings, gated subset & +22.6\% & +27.9\% \\
Mean $\Delta$VMAF, gated subset & $-2.90$ & $-4.45$ \\
Clips saving bits, gated subset & 110/122 (90.2\%) & 122/122 (100\%) \\
\bottomrule
\end{tabular}
\end{table}

The gate selects 122 of the 308 clips, or 39.6\% of the dataset. On this subset, the full-frame VMAF loss is much lower than on the full dataset. For x264, the mean change in VMAF improves from $-10.36$ in general to $-2.90$ in the gated subset. At the same time, the gated clips retain meaningful storage savings: 22.6\% with x264 and 27.9\% with x265.

This result is important for deployment. It suggests that BLUE should be used selectively rather than universally. A short pre-encode test can identify clips where the storage benefit is likely to be achieved with acceptable perceptual degradation. Since \texttt{filter\_vmaf} does not require object annotations, manual labels, or downstream detector output, it can be used in real surveillance pipelines.

\subsection{Failure Cases and Limitations Observed in Results}

The strongest failures occur when the static-background assumption does not hold. In the VIRAT 050300/050301 scene family, several clips contain dynamic or heavily textured background content. For these clips, the seed becomes a poor representation of the live background, producing large VMAF regressions. Table~\ref{tab:worstclips} lists the worst cases under x265 at CRF 28.

\begin{table}[h]
\centering
\caption{Worst-case VMAF regressions, x265 at CRF 28.}
\label{tab:worstclips}
\begin{tabular}{lccc}
\toprule
\textbf{Clip suffix} & \textbf{$\Delta$VMAF} & \textbf{Savings} & \textbf{filter\_vmaf} \\
\midrule
050300\_10 & $-50.6$ & +48.2\% & 39.2 \\
050300\_05 & $-50.6$ & +43.9\% & 39.5 \\
050301\_01 & $-47.6$ & +46.2\% & 42.5 \\
050300\_09 & $-46.9$ & +44.3\% & 43.3 \\
050300\_06 & $-46.8$ & +43.5\% & 43.3 \\
\bottomrule
\end{tabular}
\end{table}

Although these clips show large storage savings, their perceptual degradation is not acceptable. Importantly, all of them have low \texttt{filter\_vmaf} values and would be rejected by the proposed deployment gate. This supports the use of \texttt{filter\_vmaf} as a practical safeguard.

One additional failure mode is observed in VIRAT\_S\_010208\_09. This clip passes the average \texttt{filter\_vmaf} gate but shows a foreground PSNR regression of $-5.93$ dB. The likely reason is a localized or short-duration quality failure that is hidden by the average clip-level score. A more conservative gate based on the minimum \texttt{filter\_vmaf} over sliding windows would be better suited to catch this type of failure.

Another limitation appears in clips that are already highly efficient under the raw codec. Thirteen x264 clips show net storage increase rather than savings. These are near-static scenes where the baseline encoder already performs well, leaving little room for BLUE to improve compression. A secondary bypass rule based on raw bitrate, for example skipping BLUE when the raw output is below approximately 50 kbps, would avoid these cases.

\subsection{Overall Discussion}

The results show that BLUE is most useful as a selective compression tool for fixed-camera surveillance video. Across the entire data set, it provides substantial storage savings, especially with x265. It also preserves or improves foreground PSNR on most clips, which is important for surveillance scenarios where moving objects are the primary region of interest.

At the same time, BLUE introduces a full-frame perceptual penalty because it modifies background pixels. This penalty is visible in the VMAF results and should be clearly disclosed. Therefore, BLUE should not be positioned as a universal perceptual-quality improvement method. Instead, it should be positioned as a storage-reduction method that prioritizes foreground preservation and standard-codec compatibility.

The deployment gate is central to making this tradeoff practical. By using \texttt{filter\_vmaf} on a short prefix, BLUE can be activated only when the compositor-induced quality loss is expected to be small. This selective strategy retains the main storage benefit while avoiding clips where the static-background assumption fails.

\section{Conclusion}

This paper presented BLUE, a pre-encode compositor for entropy-efficient surveillance video compression. BLUE targets fixed-camera surveillance scenes in which a large fraction of the frame remains static while only a smaller foreground region contains meaningful motion. Instead of modifying the codec or producing a non-standard bitstream, BLUE modifies raw NV12 frames before encoding. It preserves live pixels in motion regions and replaces static background pixels with pixels from a persistent seed frame. This makes background regions more likely to be coded as SKIP or low-cost inter-predicted blocks by standard H.264/H.265 encoders.

The evaluation on 308 annotated VIRAT surveillance clips shows that BLUE can provide substantial storage reduction under matched codec and CRF settings. At CRF 28, BLUE reduces file size by an average of 34.6\% with x264 and 39.4\% with x265. Storage savings are observed on 95.8\% of clips for x264 and 99.4\% of clips for x265. Importantly, foreground-region PSNR, measured inside object-annotation bounding boxes, is preserved or improved on 60.7\% of the clips. This indicates that BLUE can reduce the cost of static background while maintaining the quality of surveillance-relevant moving objects.

The results also show the main limitation of the current approach. Since BLUE directly substitutes background pixels, it can reduce full-frame perceptual quality even when foreground regions remain usable. This is reflected in the VMAF results, where the full-frame perceptual score decreases by a median of 6.75 points for x264 and 8.59 points for x265 at CRF 28. Therefore, BLUE should not be interpreted as a universal improvement to full-frame rate-distortion performance. Rather, it is best understood as a selective storage-reduction method for surveillance scenarios where foreground fidelity and standard-codec compatibility are the primary requirements.

To make this tradeoff practical, the paper introduced \texttt{filter\_vmaf}, a pre-encode quality gate computed on a short prefix of the video. The gate identifies clips where BLUE is likely to provide meaningful storage reduction with acceptable perceptual degradation. With a threshold of \texttt{filter\_vmaf} $\geq$ 88, approximately 40\% of clips are selected, achieving average storage savings of 22.6\% with x264 and 27.9\% with x265, while substantially reducing the VMAF penalty. This supports a selective-deployment strategy in which BLUE is activated only when the scene characteristics are suitable.

Future work should focus on reducing the perceptual cost of background substitution and improving worst-case robustness. One direction is to use the BLUE motion mask as an encoder-side region-of-interest signal rather than directly modifying pixels. Such an approach would preserve the original frame content while guiding the encoder to allocate more bits to moving regions and fewer bits to static background. A second direction is to replace the current clip-level \texttt{filter\_vmaf} gate with a sliding-window minimum gate, which may better detect short-duration quality failures. Finally, a raw-bitrate bypass rule could avoid applying BLUE to clips where the baseline encoder is already highly efficient and little additional compression benefit is available.

Overall, BLUE demonstrates that surveillance-specific structure can be exploited before encoding to improve storage efficiency while maintaining compatibility with standard video codecs. The results suggest that pre-encode compositing, combined with selective activation, is a practical direction for reducing surveillance video storage requirements without requiring changes to existing decoding infrastructure.


\begin{thebibliography}{99}

\bibitem{robitza2017}
W. Robitza, ``CRF Guide: Constant Rate Factor in x264, x265 and libvpx,'' 2017.

\bibitem{h264}
ITU-T and ISO/IEC, ``Advanced video coding for generic audiovisual services,'' ITU-T Rec. H.264 | ISO/IEC 14496-10.

\bibitem{sullivan2012}
G. J. Sullivan, J.-R. Ohm, W.-J. Han, and T. Wiegand, ``Overview of the High Efficiency Video Coding (HEVC) Standard,'' \textit{IEEE Trans. Circuits Syst. Video Technol.}, vol. 22, no. 12, pp. 1649--1668, Dec. 2012, doi: 10.1109/TCSVT.2012.2221191.

\bibitem{watanabe2001}
H. Watanabe and K. Jinzenji, ``Sprite Coding in Object-based Video Coding Standard: MPEG-4,'' in \textit{Proc. World Multiconference on Systemics, Cybernetics and Informatics (SCI 2001)}, vol. XIII, pp. 420--425, Jul. 2001.

\bibitem{wu2020}
L. Wu, K. Huang, H. Shen, and L. Gao, ``A Foreground-background Parallel Compression with Residual Encoding for Surveillance Video,'' arXiv:2001.06590, 2020.

\bibitem{meuel2018}
H. Meuel, F. Kluger, and J. Ostermann, ``Region of Interest (ROI) Coding for Aerial Surveillance Video using AVC and HEVC,'' arXiv:1801.06442, 2018.

\bibitem{stankiewicz2024}
O. Stankiewicz et al., ``Region-of-Interest-Based Video Coding for Machines,'' in \textit{Proc. 2024 IEEE Int. Conf. Multimedia and Expo Workshops (ICMEW)}, Niagara Falls, ON, Canada, 2024, pp. 1--6, doi: 10.1109/ICMEW63481.2024.10645441.

\bibitem{kroeger2016}
T. Kroeger, R. Timofte, D. Dai, and L. Van Gool, ``Fast Optical Flow using Dense Inverse Search,'' in \textit{Proc. Eur. Conf. Comput. Vis. (ECCV)}, 2016.

\bibitem{li2016}
Z. Li, A. Aaron, I. Katsavounidis, A. Moorthy, and M. Manohara, ``Toward a Practical Perceptual Video Quality Metric,'' \textit{Netflix Technology Blog}, 2016.

\bibitem{oh2011}
S. Oh et al., ``A Large-scale Benchmark Dataset for Event Recognition in Surveillance Video,'' in \textit{Proc. IEEE Conf. Comput. Vis. Pattern Recognit. Workshops}, 2011.

\bibitem{bjontegaard2001}
G. Bj{\o}ntegaard, ``Calculation of Average PSNR Differences between RD-curves,'' ITU-T VCEG-M33, Austin, TX, USA, Apr. 2001.

\bibitem{zabih1994}
R. Zabih and J. Woodfill, ``Non-parametric Local Transforms for Computing Visual Correspondence,'' in \textit{Proc. Eur. Conf. Comput. Vis. (ECCV)}, 1994.

\end{thebibliography}
\end{document}